\documentclass[conference]{IEEEtran}
\usepackage{times,amsmath,epsfig,latexsym,amssymb,psfrag,graphicx}
\usepackage{paralist}
\usepackage{booktabs}
\usepackage{color}
 \usepackage{setspace}
\usepackage{epstopdf}
\usepackage{subcaption}
\usepackage{mathbbol}
\usepackage[ruled,vlined]{algorithm2e}
\usepackage{citesort}
\newcommand{\qed}{\nobreak \ifvmode \relax \else
\ifdim\lastskip<1.5em \hskip-\lastskip
\hskip1.5em plus0em minus0.5em \fi \nobreak
\vrule height0.75em width0.5em depth0.25em\fi}
\begin{document}

\newcommand{\CS}[1]{\textcolor{magenta}{#1}} 
\newcommand{ \PP}[1]{\textcolor{blue}{#1}} 
\newcommand{\AZ}[1]{\textcolor{green}{#1}} 

\title{Latency-Energy Tradeoff based on Channel Scheduling and Repetitions in NB-IoT Systems}

\author{Amin Azari$^*$, Guowang Miao$^*$, \v Cedomir Stefanovi\' c$^+$, and Petar Popovski$^+$\\
$^*$KTH Royal Institute of Technology, $^+$Aalborg University\\
Email: \{aazari,guowang\}@kth.se, \{cs,petarp\}@es.aau.dk}
\maketitle

\begin{abstract}

Narrowband IoT (NB-IoT) is the latest IoT connectivity solution presented by the 3GPP.
NB-IoT introduces coverage classes and introduces a significant link budget improvement by allowing repeated transmissions by nodes that experience high path loss. However, those repetitions necessarily increase the energy consumption and the latency in the whole NB-IoT system. The extent to which the whole system is affected depends on the scheduling of the uplink and downlink channels. We address this question, not treated previously, by developing a tractable model of NB-IoT access protocol operation, comprising message exchanges in random-access, control, and data channels, both in the uplink and downlink.
The model is then used to analyze the impact of channel scheduling as well as the interaction of coexisting coverage classes, through derivation of the expected latency and battery lifetime for each coverage class.
These results are subsequently employed in investigation of latency-energy tradeoff in NB-IoT channel scheduling as well as determining the optimized operation points.
Simulations results show validity of the analysis and confirm that there is a significant impact of channel scheduling on latency and lifetime performance of NB-IoT devices.
\end{abstract}

\IEEEpeerreviewmaketitle

 \section{Introduction}\label{intr}
 
Internet of Things (IoT) is behind 2 out of 3 major drivers of next generation wireless networks, which are massive IoT connectivity, mission critical IoT connectivity and enhanced mobile broadband (eMBB) \cite{5g_iot}.
Due to the fundamental differences in characteristics and service requirements between IoT and legacy traffic in cellular networks, which are seen in massive number of connected devices, short packet sizes, and long battery lifetimes, revolutionary connectivity solutions have been proposed and implemented by industry \cite{lif_com,mag_all}.
The most prominent examples of such solutions are SigFox, introduced in 2009, and LoRa, introduced in 2015, both implemented in the unlicensed band, i.e., 868 MHz in Europe \cite{mag_all,int1}.
On the other hand, the accommodation of IoT traffic over cellular networks has been investigated by the 3GPP, proposing evolutionary solutions like LTE Cat1 and LTE Cat-M  \cite{emtc,ltemm}.
Recently, these efforts have been also complemented by introduction of revolutionary solutions like NB-IoT \cite{ciot}.


 \begin{figure}[t!]
        \centering
                \includegraphics[width=\columnwidth]{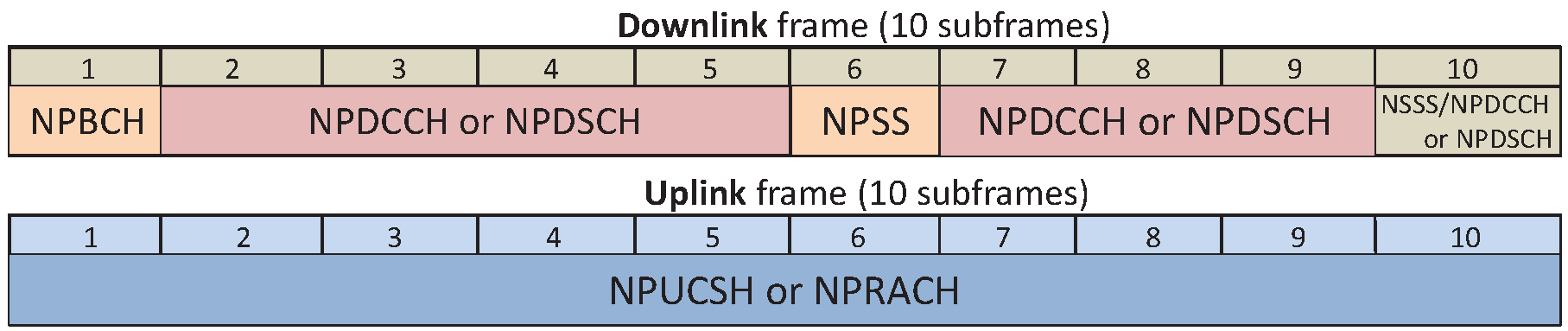}
                \caption{NB-IoT features frequency-division duplex for uplink and downlink \cite{wp}. Downlink/uplink NP channels and signals are time multiplexed, as depicted in the figure.} 
                \label{sf}
\end{figure} 

NB-IoT represents a big step towards realization of massive IoT connectivity over cellular networks \cite{nbiot}. 
Communication in NB-IoT systems takes place in a narrow, $200 \, \mathrm{KHz}$ bandwidth, resulting in more than $20 \, \mathrm{dB}$ link budget improvement over the legacy LTE.
This enables smart devices deployed in remote areas, e.g., basements, to communicate with the base station (BS). 
As the legacy signaling and communication protocols were designed for large bandwidths, NB-IoT introduces a solution with five new narrowband physical (NP) channels \cite{prim,wp}, see Fig.~\ref{sf}: random access channel (NPRACH), uplink shared channel (NPUSCH), downlink shared channel (NPDSCH), downlink control channel (NPDCCH), and broadcast channel (NPBCH).
NB-IoT also introduces four new physical signals: demodulation reference signal (DMRS) that is sent with user data on NPUSCH, narrowband reference signal (NRS), narrowband primary synchronization signal (NPSS), and narrowband secondary synchronization signal (NSSS).
Prior works on NB-IoT investigated preamble design for access reservation of devices over NPRACH \cite{nb_ra,nbiotaa}, uplink resource allocation to the connected devices \cite{nb_sch}, coverage and capacity analysis of NB-IoT systems in rural areas \cite{cell1}, coverage of NB-IoT with consideration of external interference due to deployment in guard band \cite{nbi_cov}, and impact of channel coherence time on coverage of NB-IoT systems in \cite{sasan}.
Further, in \cite{nbt}, energy consumption of IoT devices in data transmission over NB-IoT systems in normal, robust, and extreme coverage scenarios has been investigated.
The results obtained in \cite{nbt} illustrate that NB-IoT significantly reduces the energy consumption with respect to the legacy LTE, due to the existence of the deep sleep mode for the devices that are registered to the BS. 

In this paper, we address an important and so far untreated problem: when and how much resources to allocate to NPRACH, NPUSCH, NPDCCH, and NPDSCH in coexistence scenarios, where BS is serving NB-IoT devices with random activations that belong to different coverage classes.
The solution to this problem has a significant impact on the service execution and devices' performance, as the resource allocation to different channels faces inherent tradeoffs. 
The essence of the tradeoff can be explained as follows. If random access opportunities (NPRACH) occur frequently, less uplink radio resources remain for uplink data channel (NPUSCH), which increases the latency in data transmissions.
On the other hand, if NPRACH is scheduled infrequently, latency and energy consumption in access reservation increase due to the extended idle-listening time and increased collision probability.
Further, as device scheduling for uplink/downlink channels is performed over NPDCCH, infrequent scheduling of this channel may lead to wasted uplink resources in NPUSCH and increased latency in data transmissions. 
Conversely, if NPDCCH occurs frequently, the latency and energy consumption of transmissions over NPUSCH will increase.
Another important aspect studied in the paper is the impact of signal repetitions that are used by the devices that are located far away from the BS on battery lifetime and latency performance of other devices in the system.

The remainder of the paper is structured as follows.
In the next section, we outline the motivation for the development of a NB-IoT-specific analysis of channel scheduling and the reasons why the existing LTE models can not be used, and then we list the contributions of the paper.
Section~III is devoted to the system model.
Section~IV presents the analysis.
Investigation of the operational tradeoffs and performance evaluation are presented in Section~V.
Concluding remarks are given in Section~VI.

\section{Motivation and Contributions}

The literature on latency and energy analysis and optimization for LTE networks is mature \cite{4g,4ge}.
Furthermore, latency and energy tradeoffs in IoT scheduling over LTE networks were investigated in \cite{access,eem,eel}. 
However, although the NB-IoT access networking is heavily inspired by LTE, there are several crucial differences that prevent the use of the LTE models: (i) in NB-IoT, all communications happen in  a single LTE resource block, and hence, control, broadcast, random access, and data channels are multiplexed on the same radio resource, (ii) a set of coverage classes has been defined, which enables devices experiencing extreme pathloss values to become connected by leveraging on repetitions of transmitted signals, and (iii) the control plane has been adapted to IoT characteristics, enabling the devices to become disconnected for several hours while they are registered to the BS, which is not possible in LTE.
Further, the introduction of coverage classes also brings the novel concerns that are related to coexistence scenarios, where devices from different coverage classes are served within a cell and, thus, mutually impact their communication with the BS.
For example, one may consider a scenario in which the uplink is mainly occupied by random access and data transmission of devices with poor coverage, when high numbers of repetitions are required.
In such cases, the random access and data channels for other classes can not be scheduled frequently, which will affect their latency and energy performance.
In order to properly address the distinguishing features of NB-IoT, in this paper we extend the latency/energy models in \cite{sg,sg1,access,nbt}, incorporate the NB-IoT channel multiplexing, and consider coexistence of devices from a diverse set of coverage classes in the same cell. 

Specifically, the main contributions of this work are:
\begin{itemize}
\item
Derivation of a tractable analytical model of channel scheduling problem in NB-IoT systems that considers message exchanges on both downlink/uplink channels, from synchronization to service completion.
\item
Derivation of closed-form analytical expressions for service latency and energy consumption, and derivation of the expected battery lifetime model for devices connected to the network.
\item
Investigation of a latency-energy tradeoff in channel scheduling for NB-IoT systems. 
\item
Investigation of the interaction among the coverage classes coexisting in the system: performance loss in one coverage class due to an increase in number of connected devices from another coverage class.
\end{itemize}

\section{System Model}

\subsection{NB-IoT Access Networking}

Assume a NB-IoT cell with a base station located in its center, and $N$ devices uniformly distributed in it. 
In general, there are $C$ coverage classes defined in an NB-IoT cell, where the BS assigns a device to a class based on the estimated path loss between them and informs the device of its assignment.
Class $j$, $\forall j$, is characterized by the number of replicas $c_j$ that must be transmitted per original data/control packet.
For example, based on the specifications in \cite{wp}, each device belonging to group $j$ shall repeat the preamble transmitted over NPRACH $c_j\in \{1,2,4,8,16,32,64,128\}$ times. 
Further, denote by $f_j$ the fraction of devices belonging to class $j$, by $S$ the number of communication sessions that a typical IoT device performs \textit{daily} and by $p$ the probability that a device requests uplink service.
The arrival rates of uplink/downlink service requests to the system are, respectively:
\begin{align}
\label{eq:Gs}
G_u = \frac{N \, S \, p}{24 \cdot 3600}  \; \mathrm{sec}^{-1} , \; G_d = \frac{N \,S\,(1 - p)\,}{24 \cdot 3600} \; \mathrm{sec}^{-1}.
\end{align}

 \begin{figure}[t!]
        \centering
                \includegraphics[width=3.5in]{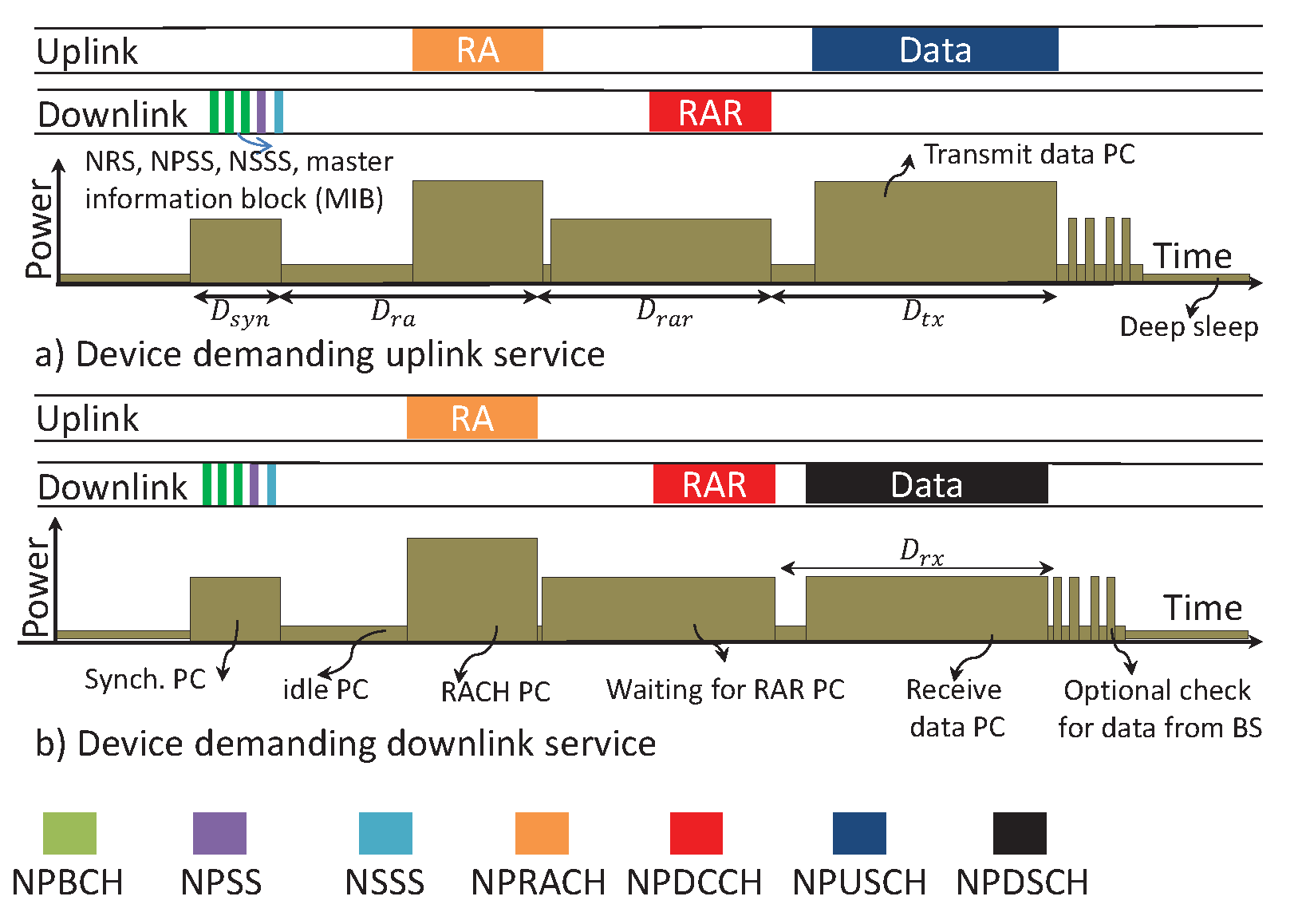}
                \caption{Communications exchanges and power consumption in NB-IoT access networking.
Note: Reference signals, including NRS, NPSS, NSSS, and master information block (MIB), are broadcasted regularly; here we show only a single  realization. 
                }
                \label{cp}
\end{figure}

Initially, when a NB-IoT device requires an uplink/downlink service, it first listens for the cell information, i.e., NPSS and NSSS, through which it synchronizes with the BS.
Then, the device performs access reservation, by sending a random access (RA) request to the BS over NPRACH.
The BS answers to a successfully received RA by sending the random access response (RAR) message over NPDCCH, indicating the resources reserved for serving the device.
Finally, the device sends/receives data to/from the BS over NPUSCH/NPDSCH channels, which, depending on the application, may be followed by an acknowledgment (ACK) \cite{wp}.
In contrast to LTE, a device that is connected to the BS can go to the \emph{deep sleep} state \cite[Section~7.3]{ciot}, which does not exist in LTE and from which the device can become reconnected just by transmitting a RA request accompanied by a random number \cite[Fig. ~7.3.4.5-1]{ciot}.
This new functionality aims to address the inefficient handling of IoT communications by LTE \cite{eel,nbt}, as it significantly saves energy due to the fact that IoT devices do not need to restart all steps of the connection establishment procedure.
Fig.~\ref{cp} represents the access protocol exchanges for NB-IoT, as described in \cite[Section~7.3]{ciot}.\footnote{For the sake of completeness, we also mention another novel reconnection scheme designed for NB-IoT, in which a device can request to resume its previous connection after receiving the random access response (RAR) \cite[Section~III]{prim}.
Towards this end, it needs to respond to the RAR message by transmission of  its previous connection ID as well as the cause for resuming the connection.}


\subsection{Problem Formulation}

Based on the model presented in Fig.~\ref{cp}, the expected latencies in uplink/downlink communication in class $j$ are, respectively:
\begin{align}
D_{{u}_j} &{=} D_{\text{sy}_j} {+} D_{\text{rr}_j} {+} D_{\text{tx}_j} \nonumber\\
D_{{d}_j} &{=} D_{\text{sy}_j} {+} D_{\text{rr}_j} {+} D_{\text{rx}_j}\label{e1}
\end{align}
where $D_{\text{sy}_j}$, $D_{\text{rr}_j}$, $D_{\text{tx}_j}$, $D_{\text{rx}_j}$  are the expected time spent in synchronization,  resource reservation, data transmission in uplink service,  and data reception in downlink service, respectively.
Similarly, the models of expected energy consumption of an uplink/downlink communication in class $j$ are:
\begin{align}
\mathcal E_{{u}_j} &  = E_{\text{sy}_j} + E_{\text{rr}_j} +  E_{\text{tx}_j} + E_s \nonumber\\
 \mathcal E_{{d}_j}  & = E_{\text{sy}_j} + E_{\text{rr}_j} + E_{\text{rx}_j} + E_s\label{e4}
\end{align}
where $E_{\text{sy}_j}$, $E_{\text{rr}_j}$, $E_{\text{tx}_j}$, $E_{\text{rx}_j}$, and $E_s$ are the expected device energy consumption in synchronization, resource reservation, data transmission in uplink service, data reception in downlink service, and optional communications like acknowledgment, respectively.
Since the energy consumption of a typical IoT device involved reporting application can be modeled as a semi-regenerative Poisson process with regeneration point at the end of each reporting period \cite{nl}, one may define the expected battery lifetime as the ratio between stored energy and energy consumption per reporting period.
In this case, the expected battery lifetime can be derived as:
\begin{align}
L_j= \frac{E_0 }{S p \mathcal E_{u_j}+ S ( 1-p ) \mathcal E_{d_j}}  \; [\mathrm{day}] \label{e5}
\end{align}
where $E_0$ is the energy storage at the device battery. 
In order to derive closed-form latency and  energy consumption  expressions, e.g., model $E_{{\text{rr}}_j}$ and $D_{{\text{rr}}_j}$,   in the sequel  we analytically investigate the performance impacts of channel scheduling, arrival traffic, and coexisting coverage classes on the performance indicators of interest.  
%
%
%

\section{Analysis}

As mentioned in Section~II, in NB-IoT systems the control, data, random access, and broadcast channels are multiplexed on the same set of radio resources. 
Thus, their mutual impact in both uplink and downlink directions are significant, which is not the case in legacy LTE due to wide set of available radio resources.
In the following, we propose a queuing model of NB-IoT access networking, which captures these interactions.

\subsection{Queuing Model of NB-IoT Access Protocol}

\begin{figure}[t!]
        \centering
                \includegraphics[width=3.5in]{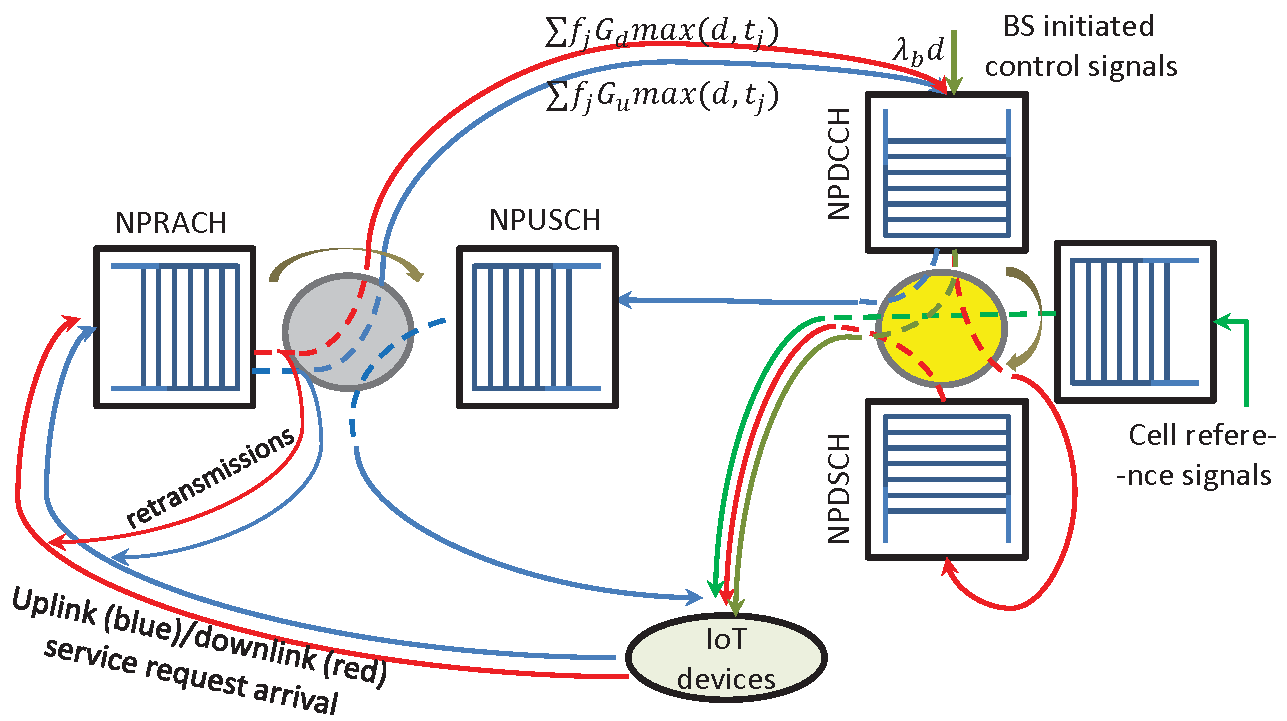}
                \caption{Queuing model of the NB-IoT access networking.
                The yellow and gray circles represent servers for downlink and uplink channels, respectively.
                }
                \label{qu}
\end{figure}

Fig.~\ref{qu} depicts the queuing model of NB-IoT access networking, 
comprising operation of NP random access, control, and data channels.
The gray circle represents the uplink server serving two channel queues, NPRACH and NPUSCH, while the yellow circle represents the downlink channel serving three channel queues, NPDCCH, NPDSCH, as well as the reference signals, such as NPSS.
Let $t_j$ be the average time interval between two consecutive scheduling of NPRACH of class $j$ and $M_j$ the number of orthogonal random access preambles available in it.
The duration of scheduled NPRACH of class $j$ is $c_j \, \tau$, where $\tau$ is the unit length, equal to the NPRACH period for the coverage class with $c_j=1$.
The inter-arrival times between two NPRACH periods in NB-IoT can vary from $40 \, \mathrm{ms}$ to $2.56 \, \mathrm{s}$ \cite{wp}.
Further, let $b$ denote the fraction of time in which reference signals are scheduled in a downlink radio frame, e.g., NPBCH, NPSS, and NSSS.
Five subframes in every two consecutive downlink frames are allocated to reference signals \cite{wp}, implying $b=0.2$.
Finally, a semi-regular scheduling of NPDCCH  has been proposed by 3GPP in order to prevent waste of resources in the uplink channel when BS serves another device with poor coverage in the downlink \cite{snpdcch}; we denote by $d$ the average time interval between two consecutive NPDCCH instances.
In the next section, we derive closed-form expressions for components of latency and battery lifetime models, given in \eqref{e1}-\eqref{e4}.

\subsection{Derivations}\label{der}

$D_{\text{sy}_j}$ in \eqref{e1} is a function of the coverage class $j$. Its average value has been reported in \cite[Sec.~7.3]{ciot}.
$D_{\text{rr}_j}$ is given by:
\begin{align} 
D_{\text{rr}_j} = \sum\nolimits_{\ell=1}^{N_{r_{\max}}} (1-\mathcal P_j)^{\ell-1}\mathcal P_j \ell ( D_{\text{ra}_j}+D_{\text{rar}_j} )
\end{align}
in which  $N_{r_{\max}}$ represents the  maximum allowed number of attempts, ${\mathcal P_j}$ the probability of successful resource reservation in an attempt that depends on the number of devices in the class attempting the access,
$D_{\text{ra}_j}$  the expected latency in sending  a RA message, and  $D_{\text{rar}_j}$ the expected latency in receiving the RAR message.
$D_{\text{ra}_j}$ is a function of time interval between consecutive scheduling of NPRACHs and is equal to $0.5 \, t_j+c_j \tau$, while $D_{\text{rar}_j}$ depends on the operation of NPDCCH.
NPDCCH can be seen as a queuing system in which the downlink server (see Fig. \ref{qu}) visits the queue every $d$ seconds and serves the existing requests.
Thus, $D_{\text{rar}_j}$ consists of i) waiting for NPDCCH to occur, which happens on average $d/2$ seconds, ii) time interval spent waiting to be served when NPDCCH occurs, denoted by $D_w$, and iii) transmission time, denoted by $D_{t_j}$.

We first characterize $D_w$. 
When the server visits the NPDCCH queue, on average there are:
\begin{align}
\mathcal Q= \sum\nolimits_{j=1}^{C}f_j(G_u+G_d)\max{\{d,t_j\}}+\lambda_b \, d
\end{align}
requests waiting to be served, where the first term in $\mathcal Q$ corresponds to  NPRACH-initiated random access requests, see \eqref{eq:Gs}, and $\lambda_b\, d$ models  the the arrival of BS-initiated control signals, see  Fig.~\ref{qu}.
Thus, the average waiting time before the service of a newly arrived RA message starts is $D_w = 0.5  \, \mathcal Q \, \mathcal D_{t}$, where $\mathcal D_t$ is the average service time in NPDCCH. 
Using $u$ as the average control packet transmission time, the average  transmission time for class $j$ is $D_{t_j}=c_j \, u$.
Thus:
\begin{align}
\mathcal D_t = \sum\nolimits_{j=1}^C f_j \mathcal D_{t_j} = \sum\nolimits_{j=1}^C f_j c_j u
\end{align}
and $D_{\text{rar}_j}$ becomes:
\begin{equation}\label{e7}
D_{\text{rar}_j}=0.5 \, d +0 .5 \, \mathcal Q \, \mathcal D_t+ c_j \, u.
\end{equation}

Resource reservation of a device over NPRACH is successful if its transmitted preamble does not collide with other nodes' preambles, which happens with probability $\mathcal P_{j_{\text{RACH}}}$, and the RA response is received within period $T_{\text{th}}$, which happens with probability $\mathcal P_{j_{\text{RAR}}}$. 
Thus, the probability of successful resource reservation can be approximated as $\mathcal P_j = \mathcal P_{j_{\text{RACH}}}\, \mathcal P_{j_{\text{RAR}}}$.
For a device belonging to class $j$, there are $M_j$ orthogonal preambles available every $t$ seconds, during which it contends on average with $\mathcal N_j=  f_j ( G_u+G_d ) t_j $ devices.
Then, $\mathcal P_{j_\text{RACH}}$ is derived as:
\begin{align}\label{e6}
\mathcal P_{j_\text{RACH}}= \sum\nolimits_{k=2}^N \frac{\mathcal (N_j)^ke^{-\mathcal N_j}}{k!} \left( \frac{ M_j - 1 }{M_j} \right)^{k-1}.
\end{align}
The cumulative distribution function of service time for a device and sum of service times for $n > 1$ devices are:
\begin{align}
\mathcal F_1(x) = \sum\nolimits_{j=1}^C f_j  H(x - c_j u), \\
\mathcal F_n(x) = \sum\nolimits_{j=1}^C f_j \mathcal F_{n - 1}(x - c_j u) \nonumber
\end{align}
respectively, where $ H(x)$ is the unit step function.
Then, $\mathcal P_{j_{\text{RAR}}}$, which is the probability that RAR is received within $T_{\text{th}}$, is:
\begin{align}
 \mathcal P_{j_{\text{RAR}}} = & 1 -  \nonumber \\
 & \sum_{K=2}^{\infty}\sum_{k=1}^{K -1}\frac{k}{K}\frac{\mathcal Q^Ke^{-\mathcal Q}}{K!} \left(1 - \mathcal F_{K - k}(T_{\text{th}}) \right) \mathcal F_{K - k - 1}(T_{\text{th}}).
\end{align}

$D_{\text{tx}_j}$ is a function of scheduling of NPUSCH.
Operation of NPUSCH can be seen as a queuing system in which server handles requests in a fraction of each uplink frame that is allocated to NPUSCH; this fraction is $w=1-\sum\nolimits_{j=1}^{C} {c_j \tau}/{t_j}$.
Arrival of service requests to the NPUSCH can be modeled as a batch Poisson process (BPP), as resource reservation happens only in NPRACH periods.
The mean batch-size is:
\begin{align}
\mathcal G= \frac{1}{C} \sum\nolimits_{j=1}^C  f_jG_ut_j
\end{align}
 and the rate of batch arrivals is $\sum\nolimits_{j=1}^C 1/t_j$. 
The uplink transmission time is determined by the packet size and coverage class $j$.
We assume that the packet length follows a general distribution with the first two moments equal to $l_{1}$ and  $l_{2}$.  
Then, the transmission (i.e., service) time for the uplink packet follows a general distribution with the first two moments:
\begin{align}
s_1 = \sum\nolimits_{j=1}^{C} \frac{f_j c_j{l_{1}}}{{\mathcal R_j}w}  \text{ and } s_2 = \sum\nolimits_{j=1}^{C} \frac{f_j c_j^2{l_{2}}}{{\mathcal R_j^2}w^2} \nonumber
\end{align}
where $\mathcal R_j$ is the average uplink transmission rate for class $j$.
This queuing system is a BPP/G/1 system, hence, using the results from \cite{booktt}, one can derive the latency in data transmission for class $j$ as:
\begin{equation}\label{e8}
D_{\text{tx}_j}=  \frac{\rho s_2}{2 s_1(1 - \rho)} + \frac{ \mathcal G s_1 }{2 (1 - \rho)} + \frac{c_j l_1}{\mathcal R_jw}
\end{equation}
where $\rho=\sum\nolimits_{j=1}^C  \mathcal G s_1/t_j$.
Similarly, performance of NPDSCH can be seen as a queuing system in which server visits the queue in a fraction of frame time and serves the requests.
This fraction comprises to subframes in which NPDCCH, NPBCH, NPSS, and NSSS are not scheduled, and can be derived similarly to \eqref{e7} as:
\begin{align}
 y = 1 -  b  - \frac{\mathcal Q}{d} \, {\sum\nolimits_{j=1}^C f_j c_j u}.
 \end{align}
The arrival of downlink service requests to the NPDSCH queue can be also seen as a BPP, as they arrive only after NPRACH has occurred.
The mean batch-size is:
\begin{align}
\mathbb G = \frac{1}{C} \sum\nolimits_{j=1}^C  f_jG_dt_j
\end{align}
and the arrival rate is $\sum\nolimits_{j=1}^C 1/t_j$. 
The downlink transmission time is  determined by the packet size and  {coverage class $j$.
Assuming that packet length follows a general distribution with moments $ m_{1}$ and  $  m_{2}$, then first two moments of the distribution of the packet transmission time are:} 
\begin{align}
h_1 = \sum\nolimits_{j=1}^{C} \frac{f_j c_j{m_1}}{{ \mathbb R_j}y }   \text{ and }  h_2\ = \sum\nolimits_{j=1}^{C} \frac{f_j c_j^2{m_2^2}}{{\mathbb R_j^2}y^2}\nonumber
\end{align}
where $\mathbb R_j$ is the average downlink data rate for coverage class $j$.
Defining $\nu=\sum\nolimits_{j=1}^C   \frac{\mathbb G h_1}{t_j}$, the latency in data reception  $D_{\text{rx}_j}$ becomes:
\begin{equation}
D_{\text{rx}_j}= \frac{0.5\nu  h_2}{h_1(1 - \nu)}{+}\frac{ \mathbb G h_1 }{2 (1 - \nu)}{+} \frac{c_j m_1}{\mathbb R_jy}.
\end{equation}

Finally, we derive the average energy consumption of an uplink/downlink service.
Denote by $\xi$, $P_I$, $P_c$, $P_l$, and $P_{t_j}$ the power amplifier efficiency, idle power consumption, circuit power consumption of transmission, listening power consumption, and transmit power consumption for class $j$. Then,
\begin{align}
& E_{\text{sy}_j}  = P_l D_{\text{sy}_j}\\
& E_{\text{rar}_j} =P_l D_{\text{rar}_j}  \label{ee1}\\ 
& E_{\text{rr}} = \sum\nolimits_{l=1}^{N_{r_{\max}}} (1-\mathcal P_j)^{l-1} \mathcal P_j( E_{\text{ra}_j}+E_{\text{rar}_j} ) \\
& E_{\text{ra}_j} = ( D_{\text{ra}}-c_j \tau ) P_I+ c_j \tau ( P_c+\xi P_{t_j} ) \\
& E_{\text{tx}_j} =  ( D_{\text{tx}_j}- \frac{c_j l_1} { \mathcal R_j w }  ) P_I+   ( P_c+\xi P_{t_j} ) \frac{c_j l_1}{\mathcal R_jw} \\
& E_{\text{rx}_j} =  ( D_{\text{rx}_j} - \frac{c_j m_1}{\mathbb R_jy}  ) P_I+   P_l \frac{c_j m_1}{\mathbb R_jy}\label{eee}
\end{align}
from which the battery lifetime model \eqref{e5} is derived as: 
\begin{align}
L_j =  E_0 \Big( & S p [E_{\text{sy}_j} + E_{\text{rr}_j}
+ E_{\text{tx}_j} + E_s] \, + \nonumber \\
  & S ( 1-p ) \mathcal [ E_{\text{sy}_j} + E_{\text{rr}_j}
+ E_{\text{rx}_j} + E_s] \Big)^{-1} .
\end{align}


\begin{table}[t!]
 \centering \caption{Parameters for performance analysis.}\label{part}
\begin{tabular}{p{1.1 cm}p{3.2 cm}p{3.4 cm}}\\
\toprule[0.5mm]
category & parameters & values\\
\midrule[0.3mm]
Traffic&$N$, $S$, $p$ & $20000$, $0.5 \, \mathrm{h}^{-1}$, $0.8$\\
Traffic&$l_1$,  $m_1$, $T_{th}$ & $500$,  $5 \, \mathrm{Kbit}$, $2 \, \mathrm{s}$\\
Traffic&$u$, $\tau$, $\lambda_b$, $b$ & $2 \, \mathrm{ms}$, $10 \, \mathrm{ms}$, 1/{\text{CF}}, $0.2$\\
Traffic&$f_1$, $f_2$& 0.5, 0.5\\ 
Power&$P_t$, $P_c$, $P_I$, $P_l$ & $0.2$, $0.01$, $0.01$, $0.1 \, \mathrm{W}$\\
Coverage& $c_1, c_2,M_1,M_2$ & $1$, $2$, $16$, $16$ \\
Coverage&$ \mathcal R_1,\mathcal R_2,\mathbb R_1,\mathbb R_2 $& $5, 5, 15, 15  \, \mathrm{Kbit/s}$ \\
Other& $E_0$,$ D_{\text{sy}_1},D_{\text{sy}_2} $ & $1 \, \mathrm{KJ} $, $0.33 \, \mathrm{s}$, $0.66 \, \mathrm{s}$ \\
Other& Commun. frame (CF) & 10 ms \\
\bottomrule[0.5mm]
\end{tabular}
\end{table}
 
\section{Performance Evaluation}


In this section, we validate the derived expressions, highlight performance tradeoffs in channel scheduling, find optimized system operation points, and identify the mutual impact among the coexisting coverage classes.
System parameters are presented in Table~I.

\begin{figure}[t!]
        \centering
                \includegraphics[trim={1mm 1mm 3mm 4mm},clip,width=3.5in]{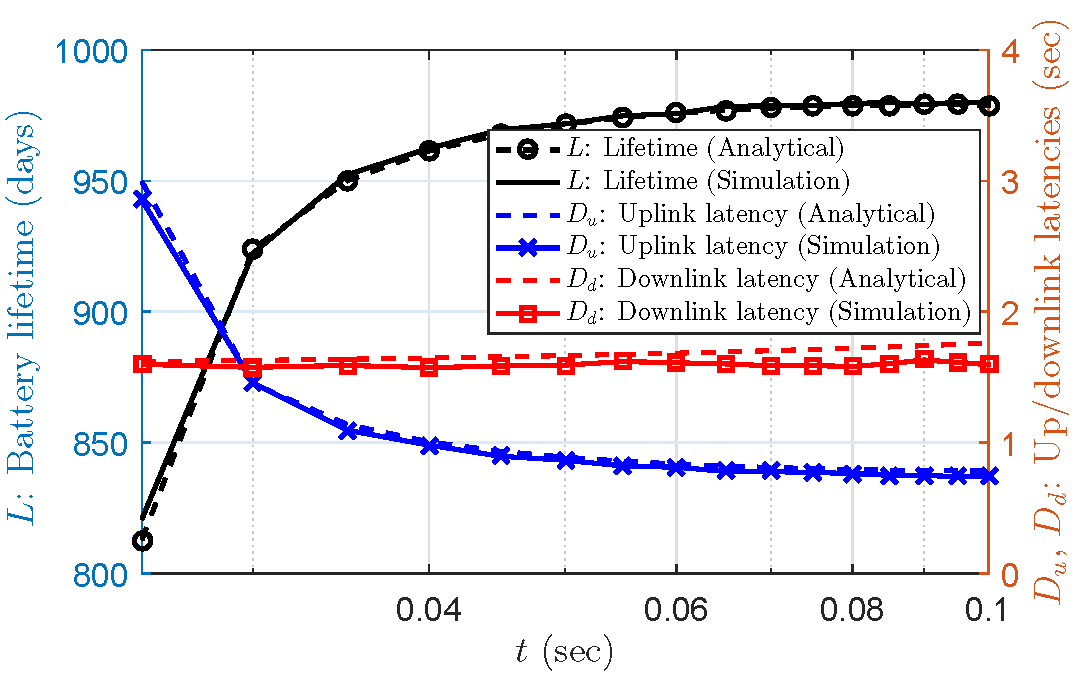}
                \caption{Comparison of analytical and simulation results versus $t$ for class 1.    $d =10 \, \mathrm{ms}$, $S=0.25\text{h}^{-1}$, $l_1=200$ bits, and $m_1=5$Kbits.}
                \label{comp}
\end{figure}

Fig.~\ref{comp} compares the analytical lifetime and latency expressions derived in Section~\ref{der} (dashed curves) against the simulation results (solid curves) for class 1 of devices. The $x$-axis represents $t$, the average time between two scheduling of random access resources.  
It obvious that the simulations results, including battery lifetime and service latency in uplink and downlink, match well with the respective analytical results.
 
\begin{figure}[t!]
        \centering
                \includegraphics[width=3.5in]{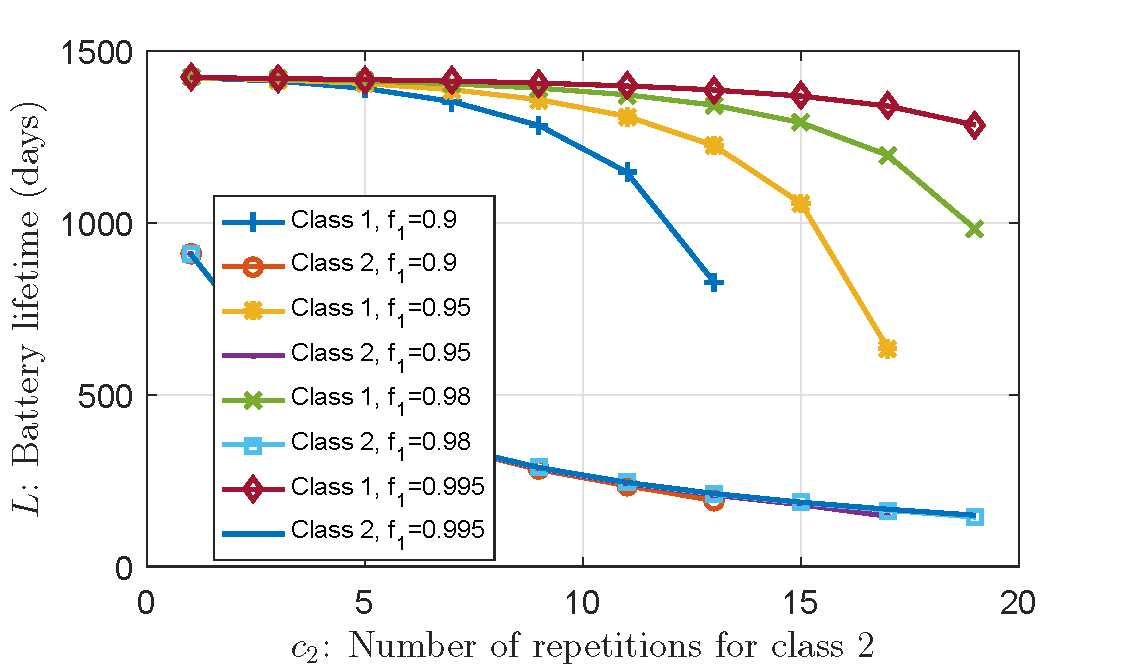}
                \caption{Mutual impact among two coexisting classes in a cell versus number of repetitions for the second class ($C = 2$, $c_1 = 1$, $f_2 =1 - f_1$, $\tau =2 \, \mathrm{ms}$, $d =10 \, \mathrm{ms}$, $t = 65 \, \mathrm{ms}$).}
                \label{mu}
\end{figure}

Fig.~\ref{mu} shows the mutual impact of two coexisting coverage classes in a cell, i.e., class~1 and class~2.
The $y$-axis represents the expected battery lifetime for both classes, while the $x$-axis represents the the number of repetitions for class 2, i.e., $c_2$.
Increase in $c_2$ increases the amount of radio resources which are used for signal repetitions (i.e., coverage extension) of devices in class 2.
This results in an increased latency both for class 1 and class 2 devices, and hence, increases the energy consumptions per reporting period and decreases the battery lifetime.
Also, it can be seen that an increase in the fraction of nodes belonging to class 2, adversely impacts the battery lifetime performance for class 1 devices.
For instance, increasing $c_2$ from 11 to 13 decreases the average battery lifetime of class 1 nodes for $6 \, \%$ when $f_1=0.95$ (i.e., $ f_2=0.05$) and for $28 \, \%$ when $f_1=0.90$ (i.e., $ f_2=0.1 $). 
Nevertheless, the extended coverage enables devices in class 2 to become connected to the BS, i.e., provides a deeper coverage to indoor areas.  

\begin{figure}[t!]
        \centering
         \begin{subfigure}[t]{0.47\textwidth}
        \centering
                \includegraphics[width=3.5in]{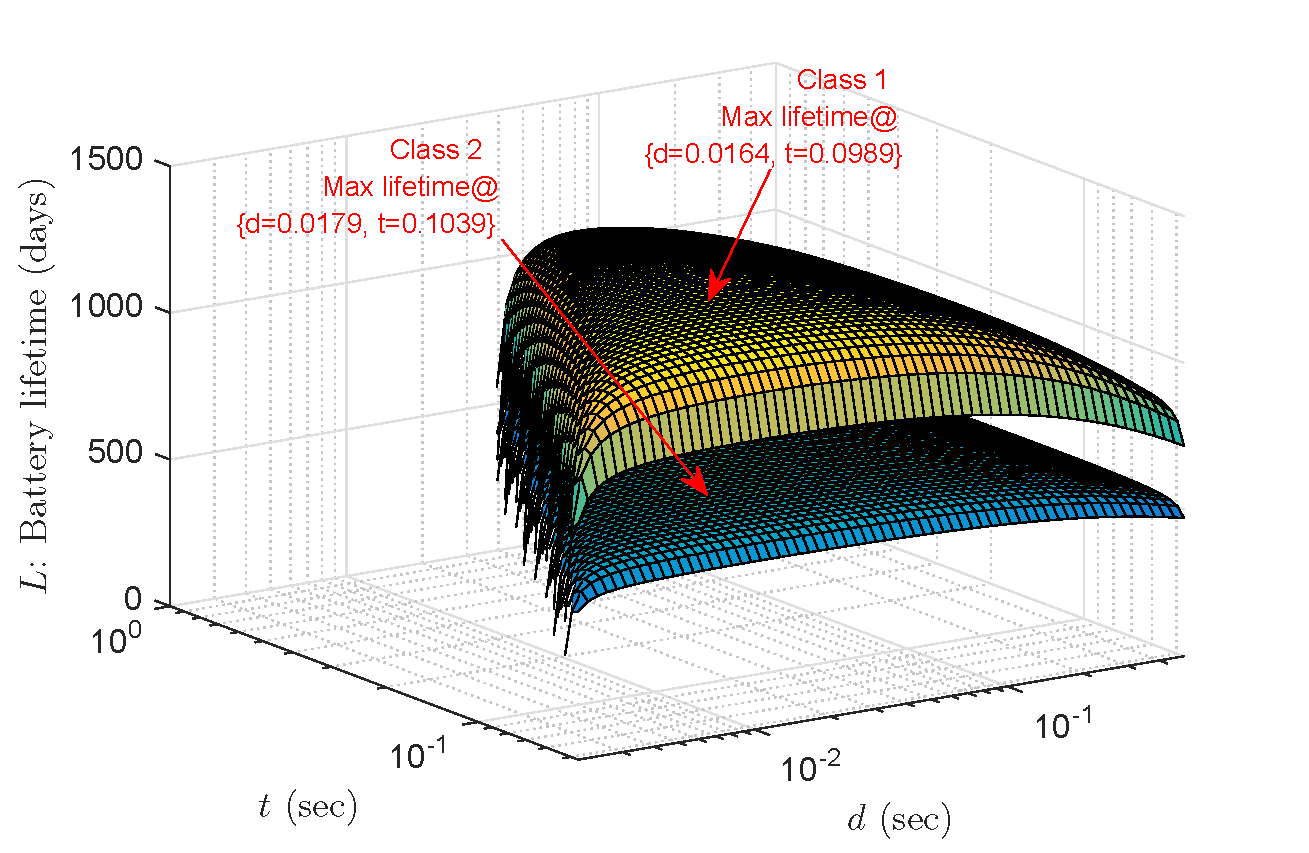}
                \caption{ Battery lifetime $L$ versus $t$ and $d$.}
                \label{bl}
\end{subfigure}\\

\begin{subfigure}[t]{0.47\textwidth}
        \centering
                \includegraphics[width=3.5in]{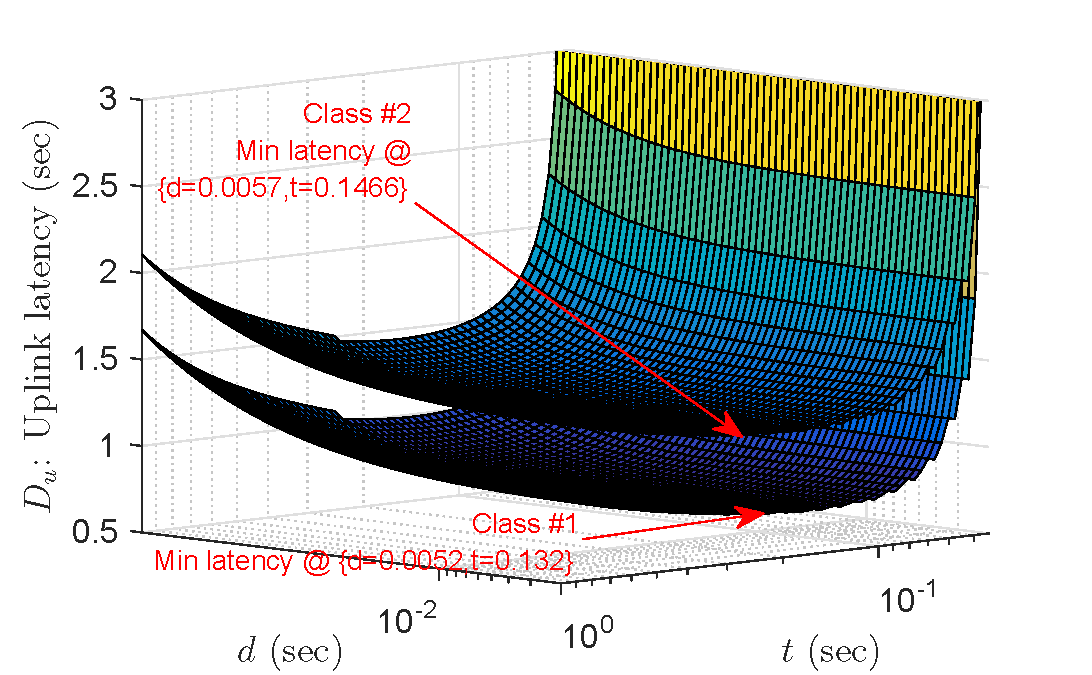}
                \caption{ Uplink latency $D_u$ versus $t$ and $d$.}
                \label{ud}
\end{subfigure}\\

\begin{subfigure}[t]{0.47\textwidth}
        \centering
                \includegraphics[width=3.5in]{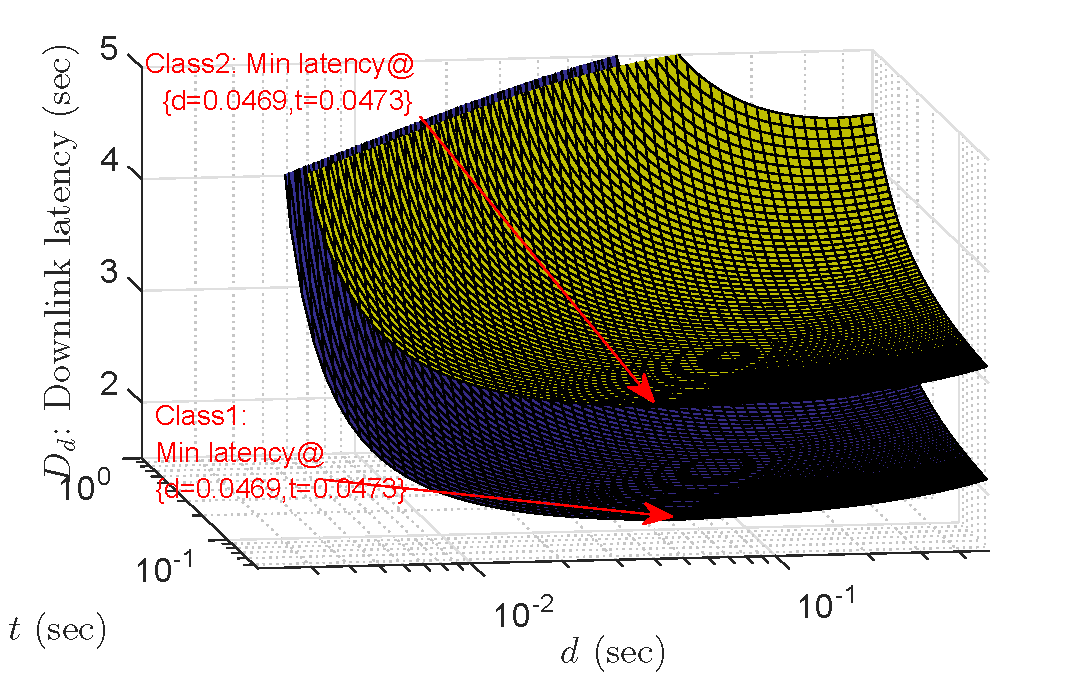}
                \caption{ Downlink latency $D_d$ versus $t$ and $d$.}
                \label{dd}
\end{subfigure}

\caption{Performance as function of $t$ and $d$, which are time intervals between two scheduling of NPRACH and NPDCCH, respectively. }\label{figam}
\vspace{-6pt}
\end{figure}

Fig.~\ref{bl} shows the expected battery lifetime versus $t$ and $d$, i.e., the time intervals between two consecutive scheduling of NPRACH and NPDCCH, respectively, for the same coexistence scenario.
Increasing $t$ at first increases the lifetime of devices in both classes, as it provides more resources for NPUSCH scheduling and decreases time spent in data transmission, i.e., $D_{tx}$.
After a certain point, increasing $t$ reduces the lifetime due to the increase of the expected time in resource reservation.
Similarly, increasing $d$ at first increases the lifetime by providing more resources for NPDSCH, decreasing the time spent in data reception, $D_{rx}$ 
while after a certain point it decreases the lifetime by increasing  the expected time in resource reservation. 

The impact of $t$ and $d$ on latency in uplink/downlink services is shown in Fig.~\ref{ud}/Fig.~\ref{dd}. 
If the uplink/downlink latency, or the battery consumption represents the only optimization objective, it is straightforward to derive the optimized operation points.
However, Figs.~\ref{bl}-\ref{dd} show that overall optimization of the objectives is coupled in conflicting ways.
This is illustrated in Fig.~\ref{opa}, which shows normalized lifetime and latency for class 1 when one of the parameters, $d$ or $t$, is fixed.
For instance, when $d=2$~ms, the downlink and uplink latency are minimized for  $t=25\, \textrm{ms}$ and $t=200\, \textrm{ms}$, and lifetime is maximized for $t=65\, \textrm{ms}$.
Also, when $t=200$~ms, the downlink and uplink latencies are minimized for  $d=200\, \textrm{ms}$ and $d=2\, \textrm{ms}$, and lifetime is maximized for $d=10\, \textrm{ms}$. 
Finally, Figs. \ref{bl}-\ref{dd} show that the latency- and lifetime-optimized resource allocation strategy differ on class basis; thus, selecting the optimized values of $t$ and $d$ depends on required quality of service (lifetime and/or latency) for each class. 


\begin{figure}[t!]
        \centering
                \includegraphics[trim={0mm 0mm 1mm 1mm},clip,width=3.5in]{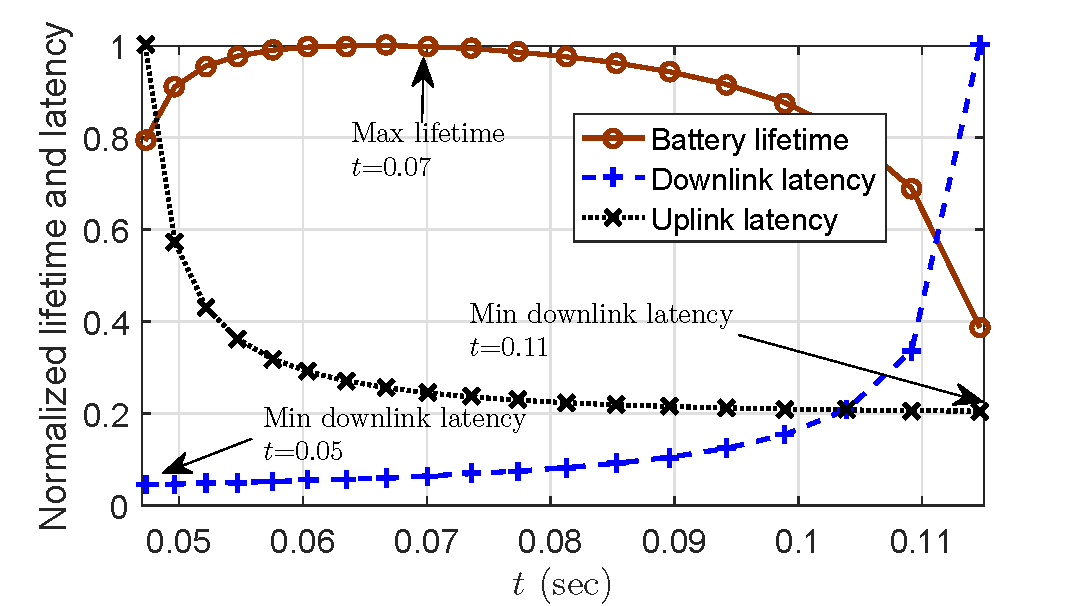}
                \caption{Overall performance analysis for class 1 vs. $t$; $d=0.0044$. }
                \label{opa}
\end{figure} 

\begin{figure}[t!]
        \centering
                \includegraphics[trim={0mm 0mm 1mm 1mm},clip,width=3.5in]{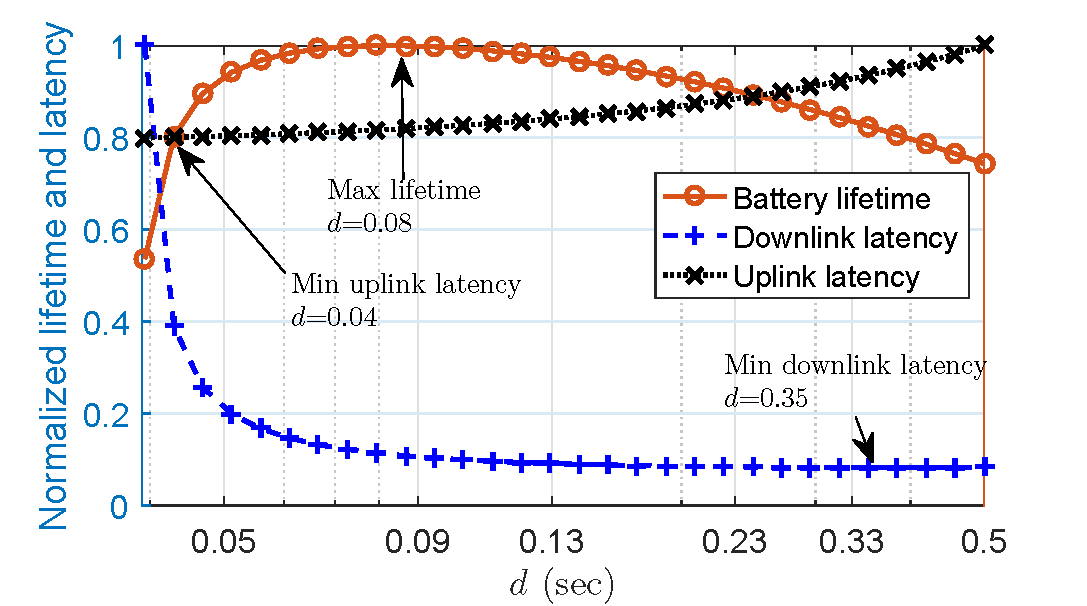}
                \caption{Overall performance analysis for class 1  vs. $d$; $t=1$.}
                \label{opa}
\end{figure}

\section{Conclusion}

NB-IoT access protocol scheduling has been investigated, and a tractable queuing model has been proposed to investigate impact of scheduling on service latency and battery lifetime.
Using derived closed-form expressions, it has been shown that scheduling of random access, control, and data channels cannot be treated separately, as the expected latencies and energy consumptions in different channels are coupled in conflicting ways.
Furthermore, the derived analytical model has been leveraged to investigate the performance impact of serving devices experiencing high pathloss, and thus needing of more signal repetitions, on latency and battery lifetime performance of other nodes.
Finally, given the set of provisioned radio resources for NB-IoT and arrival traffic, optimized scheduling policies minimizing the experienced latency and maximizing  the expected battery lifetime have been investigated.

\section*{Acknowledgment} \label{acknow}

The research presented in this paper was supported in part by Advanced Connectivity Platform for Vertical Segment (ACTIVE) and in part by the European Research Council (ERC Consolidator Grant Nr. 648382 WILLOW) within the Horizon 2020 Program.

\ifCLASSOPTIONcaptionsoff
  \newpage
\fi

\bibliographystyle{IEEEtran}
\bibliography{bibl}

\end{document}